\documentclass[twocolumn,showpacs,pra,aps,superscriptaddress]{revtex4}
\usepackage{amsmath}
\usepackage{array}
\providecommand{\openone}{\leavevmode\hbox{\small1\kern-3.8pt\normalsize1}}
\usepackage{graphicx}

\begin{document}

\title{Multi-setting Bell inequality for qudits}

\author{Se-Wan Ji}
\affiliation{Department of Physics, Korea Advanced Institute of Science and Technology, Daejeon 305-701, Korea}
\author{Jinhyoung Lee}
\affiliation{Department of Physics, Hanyang University, Seoul 133-791, Korea}
\author{James Lim}
\affiliation{Department of Physics, Hanyang University, Seoul 133-791, Korea}
\author{Koji Nagata}
\affiliation{Department of Physics, Korea Advanced Institute of Science and Technology, Daejeon 305-701, Korea}
\author{Hai-Woong Lee}
\affiliation{Department of Physics, Korea Advanced Institute of Science and Technology, Daejeon 305-701, Korea}

\begin{abstract}
  We propose a generalized Bell inequality for two three-dimensional systems
  with three settings in each local measurement. It is shown that this
  inequality is maximally violated if local measurements are configured
  to be mutually unbiased and a composite state is maximally entangled.
  This feature is similar to Clauser-Horne-Shimony-Holt inequality for
  two qubits but is in contrast with the two types of inequalities,
  Collins-Gisin-Linden-Massar-Popescu and Son-Lee-Kim, for
  high-dimensional systems.  The generalization to aribitrary
  prime-dimensional systems is discussed.
\end{abstract}


\maketitle

\section{Introduction}

Nonlocality is a profound notion in quantum mechanics. Quantitative
predictions by quantum mechanics are incompatible with constraints which
local realism implies on a correlation of measurements between two
separate systems. These constraints are called Bell inequalities \cite{Bell64}. A typical Bell inequality for bipartite two-dimensional systems (two
qubits) was derived by Clauser, Horne, Shimony, and Holt (CHSH)
\cite{CHSH69}, allowing more flexibility in local measurement
configurations than the original Bell inequality \cite{Bell64}. Quantum
mechanics maximally violates the CHSH inequality when the two qubits are
in a maximally entangled state and each qubit is measured by two
mutually unbiased bases \cite{C80,Horodecki}. We observe that nonlocality
for maximally entangled qubits is most strongly manifested by mutually unbiased
bases, similarly to the complementarity principle.

Since the discovery by Bell \cite{Bell64}, investigation of nonlocality for
more general systems has been regarded as one of the most important
challenges in quantum mechanics and quantum information science
\cite{Werner,Garg80,GHZ89,Mermin90,ABK92,KGZ00,KKCZO02,CKKO02,CGLMP02,ACG04,Lee04,Cerf02,Cabello02,SLK06,Gisin92}.
The studies include nonlocality without inequalities for three or more
qubits, presented by Greenberger, Horne, and Zeilinger \cite{GHZ89}. In
distinction with the bipartite qubit case, the contradiction between local
realism and quantum mechanics can now be revealed by perfect
correlations. Mermin immediately derived statistical inequalities for
arbitrarily many qubits and showed that the degree of their violations
exponentially increases with an increasing number of parties
\cite{Mermin90,ABK92}. The nonlocality for multipartite systems plays an
important role in quantum information processing, for instance, one way
quantum computation with cluster states \cite{RB01}.

Generalization to higher dimensional systems (qudits) has also
been investigated
\cite{KGZ00,KKCZO02,CKKO02,CGLMP02,ACG04,Lee04,Cerf02,Cabello02,SLK06}. Nonlocality of two qudits was shown to be more robust against isotropic noises than that of two qubits by numerical analysis \cite{KGZ00} and by analytically deriving Collins-Gisin-Linden-Massar-Popescu (CGLMP) inequality \cite{CGLMP02}.
Son {\it et al.} recently derived inequalities and showed their violations for arbitrary many qudits, including two qudits
\cite{SLK06}. Such inequalities for two qudits can be applied to a
bipartite division of many qubits, for instance, a division of $2n$
qubits into two parties, each having $n$ qubits, which is equivalent to a
$2^n \times 2^n$ system. We may ask when such Bell inequalities for
qudits are maximally violated: Are they maximally violated when a maximally
entangled state and mutually unbiased measurements are employed, as in the CHSH
inequality for two qubits?  It was shown that the CGLMP inequality is
maximally violated by a {\em partially entangled} state, not by any
maximally entangled states, for two three-dimensional systems (qutrits)
and further by mutually {\em biased} measurements \cite{ADGL02}. On the
other hand, the inequality of Son {\it et al.} is maximally violated by a maximally
entangled state, but still with mutually biased measurements. These
features are ``counter-intuitive'' in the sense that there exists no
nonlocality for neither entanglement nor unbiased measurements.
They are also in contrast with the CHSH inequality which is maximally violated
for a maximally entangled state and mutually unbiased measurements.

The generalized Bell inequalities mentioned above were derived by assuming that each
observer is allowed to choose one of two possible settings in the local
measurement. However, one may extend the number of measurement settings,
as done for qubits in Ref.~\cite{Z93,ZK97,NLP06}. We conjecture that the
counter-intuitive features of the generalized Bell inequalities would be
due to deficiency in the number of measurement settings, as $(d+1)$
mutually unbiased bases are possible for a prime or power-of-prime
$d$-dimensional system.

In this paper, we propose a Bell inequality for two qutrits that is
maximally violated when a maximally entangled state and mutually unbiased
measurements are employed. For the purpose we allow each observer to
choose one of three measurement settings. In addition generalization of our Bell inequality to prime-dimensional qudits is discussed.

\section{Three-setting Bell inequality for two qutrits}
\subsection{CHSH inequality for two qubits}

Before presenting Bell inequality for two qutrits, we briefly discuss the
CHSH inequality for two qubits \cite{CHSH69} as they have in common
certain properties. Suppose two parties, Alice and Bob, are
separated in a long distance and observe two qubits distributed to them.
Alice and Bob each have two sets of measuring apparatus. They each choose independently one of the two sets in their possession and perform a measurement with that set. We call the two variables, whose values are determined by the measurements using Alice's (Bob's) two sets of apparatus, $A_{0}$ and $A_{1}$ ($ B_{0} $ and $B_{1}$ ), respectively. We assign two possible values of $\pm 1$ to the outcome of the measurement on each variable. The CHSH inequality is a constraint on correlations between Alice's and Bob's measurement outcomes if local realistic description is assumed. The Bell function for
CHSH inequality is given as \cite{BMR92},
\begin{equation}
  \label{e1}
  {\cal B}(\lambda)=A_{0}(\lambda) \left( B_{0}(\lambda) +B_{1}(\lambda) \right) + A_{1}(\lambda) \left( B_{0}(\lambda) - B_{1}(\lambda) \right),
\end{equation}
where $\lambda$ is a collection of local hidden variables and the variables, $A_{i}(\lambda)$ and $B_{j}(\lambda)$, take $\pm 1$
depending on the hidden variables $\lambda$, respectively. According to the local
hidden variable theory, the statistical average of the Bell function must
satisfy the following inequality \cite{CHSH69,C80,BMR92},
\begin{equation}
  \label{e2}
  -2 \leq \langle {\cal B} \rangle \leq 2,
\end{equation}
where the statistical average $\langle {\cal B} \rangle = \int d\lambda
\rho(\lambda) {\cal B}(\lambda)$ with a probability density distribution
$\rho(\lambda)$.

Taking a quantum-mechanical description, the statistical average of
the Bell function is replaced by a quantum average of the corresponding operator \cite{CHSH69,C80,BMR92}. The Bell operator, the
counterpart to the classical Bell function of Eq.~(\ref{e1}), is given
as
\begin{equation}
  \label{e3}
  \hat{\cal B}=\hat A_{0} \otimes \left( \hat B_{0} + \hat B_{1} \right) + \hat A_{1} \otimes \left( \hat B_{0} - \hat B_{1} \right),
\end{equation}
where $\hat{A}_i$ and $\hat{B}_j$ are operators corresponding
to the variables $A_i$ and $B_j$, respectively. As measurement outcomes are assumed to
be $\pm 1$, each of the operators $\hat A_{i}$ and $\hat
B_{j}$ has eigenvalues $\pm 1$.

A quantum expectation of the Bell operator $\hat{\cal B}$ can be shown
to violate the CHSH inequality (\ref{e2}). Let the operators be
\begin{equation}
  \label{e4}
  \hat A_{0} = {\hat \sigma}_x,\; \hat A_{1} = {\hat \sigma}_y, \; \hat B_{0}={\hat \sigma}_x,\;\hat B_{1}={\hat \sigma}_y,
\end{equation}
where $\hat{\sigma}_{x,y}$ are Pauli operators. Further let the two qubits be in a maximally entangled state,
\begin{equation}
  \label{e5}
  |\psi \rangle =\frac{1}{\sqrt{2}} \left( |00\rangle + (-1)^{1/4}|11\rangle \right),
\end{equation}
where $\{|j\rangle \} \equiv \{ |0\rangle , |1\rangle \}$ is a standard basis whose elements are
eigenvectors of Pauli operator $\hat{\sigma}_z$. A straightforward algebraic calculation shows that the quantum
expectation $\langle \psi | \hat{\cal B} | \psi \rangle$ is $2 \sqrt{2}$
and violates the constraint of the CHSH inequality~(\ref{e2}). This implies
that any local hidden variable theories can not simulate the
quantum-mechanical correlation.

For the two-qubit nonlocality, we would remark that a) each observer
randomly chooses one of {\em two} possible settings in measuring his/her
qubit, b) each measurement produces one of two possible outcomes $\pm
1$, and c) a quantum expectation can {\em maximally} violate the
constraint, imposed by local realistic description, and reaches the
quantum maximum $2\sqrt{2}$ if two conditions of a quantum
state being maximally entangled and two local operators being
mutually {\em unbiased} are satisfied \cite{C80,Horodecki}. 

\subsection{Derivation of the three-setting Bell inequality for two qutrits}

Now we derive a three-setting Bell inequality for two qutrits. Our
derivation is motivated by the fact that Bell inequalities for
high-dimensional systems, suggested in literatures, are {\em maximally}
violated only when local operators are mutually {\em biased}
and/or a quantum state is {\em partially} entangled, contrary to the CHSH
inequality for two qubits \cite{BMR92,CGLMP02,ACG04,SLK06}. Alice and Bob now have three sets of measuring apparatus each, from which they each choose one and perform a measurement. The three variables whose values are determined by the measurements using Alice's (Bob's) three sets are referred to as $A_0$, $A_1$, and $A_2$ ($B_0$, $B_1$, and $B_2$), respectively. We assign three possible values of $1$, $\omega$, and $\omega ^{2}$, where $\omega=\exp{(i 2\pi/3)}$ is a primitive third root of unity, to the outcome of the measurement on each variable. As discussed for the CHSH inequality, the local realistic description implies that the values of the variables are predetermined by the local hidden variables $\lambda$: $A_i = A_i(\lambda)$ and $B_j = B_j(\lambda)$, and a
statistical average of their correlations is given as
\begin{equation}
  \label{eq1}
  \langle A_{i} B_{j} \rangle = \int {d\lambda \rho(\lambda)} A_i(\lambda) B_j(\lambda),
\end{equation}
where $\rho(\lambda)$ is the probability density distribution over
$\lambda$: $\rho(\lambda) \geq 0$ and $\int {d\lambda \rho(\lambda)
  =1}$.

To derive a constraint for the classical correlations, consider the
following Bell function,
\begin{equation}
  \label{eq3}
  {\cal B}(\lambda) = \frac{1}{2}\sum_{n = 1}^2 {\sum_{i = 0}^2
    {\sum_{j = 0}^2 {\omega ^{nij} A_i^n(\lambda)  B_j^n(\lambda) } } },
\end{equation}
where $A_i^n$ ($B_j^n$) is the $n$-th power of $A_i$($B_j$). This Bell
function has notable features: First, it contains higher-order
correlations, while the CHSH inequality involves only the first-order
correlations. In fact the second power of a dichotomic variable in the CHSH inequality is meaningless as it is just unity. On the other hand,
the variables contained in Eq.~(\ref{eq3}) are trichotomic variables and thus their second powers
have their own significance. Second, ${\cal B}(\lambda)$ has Bob's (or
Alice's) variables in the form of Fourier transformation. In this
perspective one may look at the CHSH inequality in the similar form and in this
sense the Bell function in Eq.~(\ref{eq3}) generalizes CHSH to qutrits.

We find classical upper and lower bounds for the statistical average of
the Bell function in Eq.~(\ref{eq3}). Note first that every statistical
average of ${\cal B}(\lambda)$ satisfies,
\begin{equation}
  \label{eq2}
  \min_\lambda  {\cal B}(\lambda) \leq \int {d\lambda \rho \left( \lambda  \right)} {\cal B}(\lambda)\leq \max_\lambda  {\cal B}(\lambda),
\end{equation}
where $\min_\lambda {\cal B}(\lambda)$ ($\max_\lambda {\cal
  B}(\lambda)$) means a minimum (maximum) of ${\cal B}$ over $\lambda$.
This is clear due to the fact that $\rho(\lambda)$ is a probability
density distribution: $\rho(\lambda) \geq 0$ and $\int {d\lambda
  \rho(\lambda) =1}$.  The classical upper and lower bounds are thus
determined by finding the maximum and minimum of the Bell function
${\cal B}(\lambda)$ over $\lambda$. By definition, each variable takes
an element in $\{1=\omega^0,\omega,\omega^2\}$ so that $A_i(\lambda) =
\omega^{a_i(\lambda)}$ and $B_j(\lambda) = \omega^{b_j(\lambda)}$ for
some integer-valued functions $a_i(\lambda)$ and $b_j(\lambda)$ with
respective to $\lambda$. Then Eq.~(\ref{eq3}) can be rewritten as
\begin{eqnarray}
  \label{eq4}
  {\cal B}(\lambda) &=& \frac{1}{2} \sum_{n = 1}^2 \sum_{i = 0}^2
  \sum_{j = 0}^2 \left[ {\omega^{n (a_i(\lambda)  +
        b_j(\lambda) + I j)} } \right] \nonumber \\
  &=& \frac{3}{2}\left(
    \sum_{i = 0}^2 \sum_{j = 0}^2 \delta (
    a_i(\lambda)  + b_j(\lambda)  + i j) - 3 \right),
\end{eqnarray}
where $\delta (a)=1$ if $a \equiv 0 \bmod{3}$ and $\delta (a)=0$
otherwise. Here, we used the identity, $\sum_{n=0}^{2}{\omega^{a
    n}}=3\delta (a)$.  Determining the upper and lower bounds of the
Bell function ${\cal B}(\lambda)$ reduces to finding the bounds of
$\Delta={\sum_{i,j} {\delta( {a_i + b_j + ij} )} }$ over arbitrary
integers $a_i$ and $b_j$ modulo 3.

Meanwhile, we present two useful facts resulting from a number theory
(see Ref.~\cite{AG76}). First, for a given prime integer $d$, $Z_d =
\left\{ {0,1,...,d - 1} \right\}$ is a complete set of residues modulo
$d$ so that $a Z_d \equiv \{ 0 a, 1a, ..., (d-1)a \} = Z_d$ for an
arbitrary integer $a \ne 0$. For instance let $d=3$ and $a=2$. Then $a
Z_{3} = \{0 a, 1 a, 2 a\} = \{0, 2, 1\} = Z_3$. Second, for $a, b, c \in
Z_d$, $a b \ne a c \bmod{d}$ if and only if $b \ne c \bmod{d}$.

Returning to the problem of finding the bounds of $\Delta$, consider a matrix with elements consisting of the arguments of the delta function in $\Delta$,
\begin{equation}
  \label{eq5}
  \begin{pmatrix}
    a_0  + b_0 & a_0  + b_1 & a_0  + b_2 \\
    a_1  + b_0 & a_1  + b_1  + 1 & a_1  + b_2  + 2 \\
    a_2 + b_0 & a_2 + b_1 + 2 & a_2 + b_2 + 1
  \end{pmatrix}.
\end{equation}
The maximum of $\Delta$, $\Delta_{\mathrm{max}}$, is decided by counting
the number of matrix elements that can simultaneously be congruent to
zero modulo 3. Suppose that two different elements in $i$-th row are
both congruent to zero modulo $3$: For $j \ne k$,
\begin{equation}
  \label{eq6}
  a_i  + b_j  + ij = a_i  + b_k  + ik = 0 \mod{3}.
\end{equation}
This is followed by
\begin{equation}
  \label{eq7}
  (b_j  - b_k) + i\left( {j - k} \right) = 0  \mod{3}.
\end{equation}
Then, the two elements in $l(\ne i)$-th row, $a_l + b_j + lj$ and $a_l +
b_k + lk$ can not simultaneously be congruent to zero modulo 3. That is,
\begin{equation}
  (a_l + b_j + lj) - (a_l + b_k + lk) = (b_j - b_k) + l (j-k) \ne 0,
\end{equation}
which results from Eq.~(\ref{eq7}) by noting $i\left( {j - k} \right)
\ne l\left( {j - k} \right) \bmod{3}$ for $i \ne l \bmod{3}$. Similar
conditions are also derived for columns. Under the conditions, consider
a case in which all the elements at the first row are zero and then one
element at the second or third row can be zero, resulting in $\Delta=5$.
Consider another case in which the first two elements at the first row
are zero and then one of the first two elements at the second or third
row can be zero as well as the last element at the second or third row,
resulting in $\Delta=6$. All other cases are equivalent to the two cases
discussed. We thus obtain $\Delta_{\mathrm{max}}=6$, for instance, when
$\{a_{0}=0, a_{1}=0, a_{2}=1, b_{0}=0, b_{1}=0, b_{2}=1 \}$. The minimum
of $\Delta$, $\Delta_{\mathrm{min}}=0$, is easily obtained by noting
$\Delta \ge 0$ and $\Delta = 0$ when $\{a_{0}=0, a_{1}=0, a_{2}=1,
b_{0}=1, b_{1}=1, b_{2}=2 \}$. The two bounds, $\Delta_{\mathrm{max}}$
and $\Delta_{\mathrm{min}}$ imply that the Bell function satisfies the
following inequality,
\begin{equation}
  \label{eq9}
  - \frac{9}{2}\,\leq {\cal B}(\lambda)\,\leq\frac{9}{2}.
\end{equation}
From both inequalities~$(\ref{eq2})$ and $(\ref{eq9})$, therefore, every
statistical average of ${\cal B}(\lambda)$ satisfies
\begin{equation}
  \label{eq10}
  - \frac{9}{2}\,\leq \langle {\cal B} \rangle\,\leq\frac{9}{2}.
\end{equation}

\subsection{Quantum violation of the three-setting Bell inequality for two qutrits}

We now show that a quantum expectation violates the Bell
inequality~(\ref{eq10}). The Bell operator corresponding to the
classical Bell function in Eq.~(\ref{eq3}) is given as
\begin{equation}
  \label{eq11}
  \hat {\cal B}  = \frac{1}{2}\sum_{n = 1}^2 {\sum_{i = 0}^2
    {\sum_{j = 0}^2 { \omega ^{nij} \hat A_i^n \otimes  \hat B_j^n } } }.
\end{equation}
Here, each operator $\hat{A}_i$ ($\hat{B}_j$) represents a measurement
for $A_i$ ($B_j$) on Alice's (Bob's) qutrit. An {\em orthogonal}
measurement of $M \in \{A_i, B_j\}$ is described by a complete set of
orthonormal basis vectors $\{|k \rangle_{M}\}$. Distinguishing the
measurement outcomes is indicated by a set of eigenvalues. Let the set
of eigenvalues be $\{1, \omega, \omega^2\}$, as the trichotomic
variable $M$ takes an element in the set by definition. The
measurement operator is then represented by $\hat{M} = \sum_{k=0}^{2}
\omega^k |k\rangle_{M M} \langle k|$.  In this representation each
trichotomic operator $\hat{M} \in \{\hat{A}_i,
\hat{B}_j\}$ is unitary, satisfying ${\hat M}^3 = \openone$ where
$\openone$ is the identity operator \cite{Cerf02,Lee04,SLK06}. We note
that the unitary operator $\hat{M}$ and its second power $\hat{M}^2$
have the same measurement basis just with different orderings of
eigenvalues so that the introduction of higher powers does not alter the
number of measurement settings in this work.

To see the quantum-mechanical violation, consider the following
unitary operators,
\begin{eqnarray}
  \label{eq13}
  \begin{matrix}
    \hat{A}_0 = \hat{f}_{10}, & \hat{A}_1 = \omega^2 \hat{f}_{11}, &
    \hat{A}_2
    = \hat{f}_{12}, \\
    \hat{B}_0 = \hat{f}_{10}, & \hat{B}_1 = \hat{f}_{12}, & \hat{B}_2 =
    \omega^2 \hat{f}_{11},
  \end{matrix}
 \end{eqnarray}
where $\{\hat{f}_{ij} \equiv \hat{X}^i \hat{Z}^j\}$ forms an orthogonal
basis on the Hilbert-Schmidt space of operators such that
$\mathrm{Tr}{\hat{f}_{ij}^\dagger \hat{f}_{kl}} = 3 \delta_{ik}
\delta_{jl}$ and each $\hat{f}_{ij}$ is a trichotomic operator with
eigenvalues $1$, $\omega$, and $\omega^2$. [It is known that every pair
of operators in $\{\hat{f}_{01}, \hat{f}_{10}, \hat{f}_{11},
\hat{f}_{12}\}$ is mutually unbiased \cite{Wootters,BBRV02}.] The
operators $\hat{X}$ and $\hat{Z}$ are $3$-dimensional Pauli operators
\cite{G99} such that
\begin{equation}
  \hat{X}|k\rangle = |k+1\rangle,\; \hat{Z}|k\rangle = \omega^{k}|k\rangle,\; \hat{X}^{3}= \openone, \hat{Z}^{3}= \openone, \nonumber
\end{equation}
where $\{|k\rangle\}$ is a standard orthonormal basis consisting of
eigenstates of $\hat{Z}$.  Consider further a maximally entangled state
of qutrits,
\begin{equation}
  \label{eq14}
  \left| \psi  \right\rangle  = \openone \otimes \hat{P} |\psi_0\rangle,
\end{equation}
where $|\psi_0\rangle = \sum_k |kk\rangle/\sqrt{3}$ and a phase shifter
$\hat{P} = \sum_k \omega^{-k/3}|k \rangle\langle k|$.

\begin{figure}[b]
\includegraphics[width=0.45\textwidth]{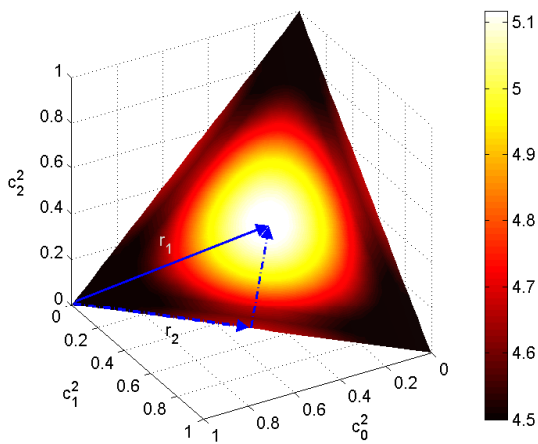}
\caption{(color online) Maximum of the quantum Bell function for each quantum state in the form of Eq.~(\ref{eq16}), which we numerically obtain over all possible operators. Quantum states are denoted by points on the triangle, defined by the plane of $\sum_i c_i^2=1$ with $c_i^2$ being the Schmidt coefficients. The vertices represent products states of Schmidt rank 1, the points on the edges 2d entangled states of rank 2, and the interior points 3d entangled states of rank 3. It is evident that the quantum Bell function reaches its maximum over all possible quantum states if the state is 3d maximally entangled with $c_0^2=c_1^2=c_2^2=1/3$.}
\label{fig1}
\end{figure}

By using the unitary operators in Eq.~(\ref{eq13}) and the maximally
entangled state in Eq.~(\ref{eq14}), the quantum expectation of the Bell
operator $\hat{\cal B}$ is given as
\begin{eqnarray}
  \label{eq15}
  \langle \psi |{\hat{\cal B}}| \psi \rangle &=&  \langle \psi_0 | \left(\openone \otimes \hat{P}^\dagger \right) {\hat{\cal B}} \left(\openone \otimes \hat{P}\right) | \psi_0 \rangle + \mathrm{c.c.} \nonumber \\
  &=& \langle \psi_0 | \left(\frac{\sqrt{3}}{2} \, \omega^{1/12} \, \sum_{i=0}^2 \hat{f}_{1i} \otimes \hat{f}_{1,-i} \right) | \psi_0 \rangle + \mathrm{c.c.} \nonumber \\
  &=& 3\sqrt 3 \cos \left( {\frac{\pi }{{18}}} \right) \approx 5.117,
\end{eqnarray}
where c.c. stands for the complex conjugate and the subscripts $i$ and
$j$ in $\hat{f}_{ij}$ are congruent to positive residues modulo 3. In
Eq.~(\ref{eq15}) we sequentially used two facts: a) The phase shifter
$\hat{P}$ transforms Bob's operators according to
\begin{eqnarray}
  \hat{P}^\dagger \hat{B}_{i} \hat{P} &=& \frac{\omega^{1/12}}{\sqrt{3}}
  \sum_{j=0}^2 \omega^{(i - j + 1)j} \hat{f}_{1j}.
\end{eqnarray}
b) The maximally entangled state $|\psi_0\rangle$ is a common eigenstate
of three composite operators, that is, $\hat{f}_{1i} \otimes
\hat{f}_{1,-i} | \psi_0 \rangle = |\psi_0 \rangle$ for all $i=0,1,2$,
implying the perfect correlations for these composite variables.
Then, the quantum expectation in Eq.~(\ref{eq15}), $3\sqrt{3}
\cos(\pi/18)\approx 5.117$ clearly exceeds the classical upper bound
$9/2=4.5$ of Bell inequality~(\ref{eq10}). This shows the nonlocality
for two qutrits with three settings of local measurements by each
observer. 

\begin{figure}[b]
\includegraphics[width=0.45\textwidth]{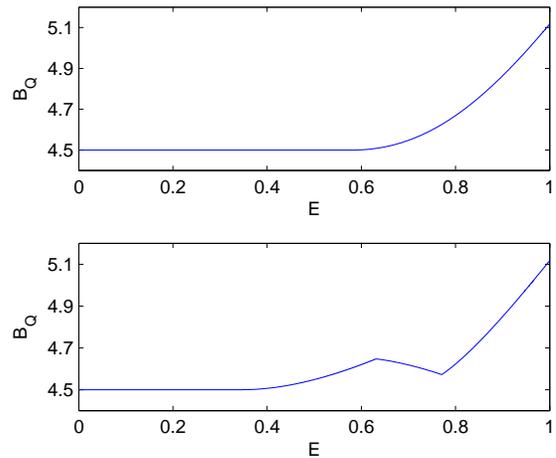}
\caption{Maximum of the quantum Bell function $B_Q$ with respect to the degree of entanglement $E$ for quantum states on the routes (a) $r_1$ and (b) $r_2$, shown in Fig.~1, from the product state $|00\rangle$ to the 3d maximally entangled state $(|00\rangle+|11\rangle+|22\rangle)/\sqrt{3}$. The route $r_1$ includes 3d entangled states, as in Eq.~(\ref{eq16}), with $c_2=c_1$ and $c_1 \le c_0$. The route $r_2$ includes 2d entangled states with $c_2=0$ and then 3d entangled states with $c_0=c_1$ and $c_1 \ge c_2$. The global maximum is achieved for the 3d maximally entangled state in both cases.}
\label{fig2}
\end{figure}
\subsection{Maximal violation of the three-setting Bell inequality}

We investigate if the quantum expectation in Eq.~(\ref{eq15}) is maximal over all possible states. For the purpose it is necessary to optimize the quantum Bell function over all possible operators for each entangled state. By employing steepest decent method (see Ref.~\cite{Son04} for the detailed methodology), we numerically find a set of such optimal unitary operators ($\hat{A}_i$ and $\hat{B}_j$) under local unitary transformations of SU(3). A pure state can in general be written, by Schmidt decomposition, as
\begin{eqnarray}
\label{eq16}
|\psi' \rangle = c_0 |00 \rangle + c_1 |11\rangle + c_2 |22\rangle,
\end{eqnarray}
where $c_i$ are non-negative real numbers, satisfying $\sum_i c_i^2=1$. In Fig.~1, composite states of two qutrits are denoted by points on the triangle, defined by the plane of $\sum_i c_i^2 = 1$ in the three dimensional vector space with the axes being Schmidt coefficients $c_i^2$. The vertices represent products states of Schmidt rank 1, the points on the edges two-dimensional (2d) entangled states of rank 2, and the interior points three-dimensional (3d) entangled states of rank 3.
Fig.~1 presents the maximum of the quantum Bell function for a given quantum state $|\psi'\rangle$, which we numerically obtain over all possible operators. It clearly shows that the quantum Bell function reaches its maximum value given in Eq.~(\ref{eq15}) over all possible quantum states if the state is 3d maximally entangled with $c_0^2=c_1^2=c_2^2=1/3$.

More explicitly, we consider quantum states on two routes $r_1$ and $r_2$, shown in Fig.~1, from the product state $|00\rangle$ to the 3d maximally entangled state $(|00\rangle+|11\rangle+|22\rangle)/\sqrt{3}$. These routes are chosen due to the three-fold rotational and reflectional symmetries of the quantum-state triangle under SU(3) transformations. Fig.~2 presents the maximum of the quantum Bell function $B_Q$ with respect to the degree of entanglement $E$ for quantum states on the routes (a) $r_1$ and (b) $r_2$, where $E = - \mbox{Tr} \hat{\rho} \log_3 \hat{\rho}$ with $\hat{\rho}$ a marginal density operator. The route $r_1$ includes 3d entangled states, as in Eq.~(\ref{eq16}), with $c_2=c_1$ and $c_0 \ge c_1$. It is clearly seen in Fig.~2(a) that, as the degree of entanglement $E$ is increased, $B_Q$ monotonically increases and reaches its maximum in Eq.~(\ref{eq15}) for the 3d maximally entangled state. The route $r_2$ includes 2d entangled states with $c_2=0$ and then 3d entangled states with $c_0=c_1$ and $c_1 \ge c_2$. From Fig.~2(b), as increasing $E$, $B_Q$ increases to the local maximum when the quantum state is 2d maximally entangled, decreases slightly, and increases again to the global maximum in Eq.~(\ref{eq15}) when the state is 3d maximally entangled. Thus, {\em it is evident that our quantum Bell function reaches its maximum in Eq.~(\ref{eq15}) only if a quantum state is 3d maximally entangled as in Eq.~(\ref{eq14}).} It is worth noting that a partially entangled state results in the local maximum in our quantum Bell function, whereas CGLMP quantum Bell function admits the global maximum for a partially entangled state \cite{ADGL02}. In a sense our Bell inequality is free of the problem that the CGLMP Bell function has.

We remark that our Bell inequality is maximally violated by quantum
mechanics if a composite state is maximally entangled {\em and} the
local measurements are mutually unbiased as in
Eqs.~(\ref{eq13}) and (\ref{eq14}). Two measurements are said to be mutually unbiased if precise knowledge in one of them
implies that all possible outcomes in the other are equally probable
\cite{Lee03,Englert92}. Consider a nondegenerate and orthogonal
measurement $M$ represented by a basis $\{|k\rangle_M\}$.  Suppose a
quantum system in $d$-dimensional Hilbert space is prepared in such a
state that the outcome in the measurement $M$ can be predicted with
certainty, for instance, the system's state is $|i\rangle_M$. Let $N$ be
another nondegenerate and orthogonal measurement represented by a basis
$\{|j\rangle_N\}$. The measurement $N$ is mutually unbiased to $M$
if outcomes of measurement $N$ are equally probable for each
$|i\rangle_M$:
\begin{eqnarray}
  \label{eq:cmcm}
  p_{j|i} \equiv \left|_N\langle j|i\rangle_M \right|^2 = \frac{1}{d}, ~~~~~~\forall j=1,2,..,d.
\end{eqnarray}
The two measurement bases, $\{|i\rangle_M\}$ and
$\{|j\rangle_N\}$, are then said to be mutually unbiased.  The eigenstates of
$\hat{A}_i$ ($\hat{B}_j$) in Eq.~(\ref{eq13}) are easily determined by
noting that the eigenstates $\{|k\rangle_i\}$ of $\hat{f}_{1i}$ are
given as
\begin{eqnarray}
  |k\rangle_i &=& \frac{1}{\sqrt{3}} \sum_{l=0}^2 \omega^{-i l^2 - k l} |l\rangle.
\end{eqnarray}
It was shown that two bases $\{|k\rangle_i\}$ and $\{|l\rangle_j\}$ are
mutually unbiased if $i \ne j$ \cite{Wootters}. The unitary operators
$\hat{A}_i$ and $\hat{B}_j$ have the same bases as their corresponding
$\hat{f}$'s in Eq.~(\ref{eq13}) with different orderings of eigenvalues
so that arbitrary two local measurements represented by $ \{ \hat{A}_i \} $ or $ \{ \hat{B}_j \} $ are mutually unbiased. \\
\\

We wish to remark here on the previous work by Buhrman and Massar \cite{BM05}, in which the authors introduced a Bell function and determined its quantum upper bound allowed for the general case of $d$-dimensional systems and $d$ measurement settings when local measurements on quantum entangled states are made. The quantum upper bound they determined is "non-tight" in the sense that their Bell function cannot take on a value greater than that, but it has not been proven that this upper bound can actually be attained. Applying their result to our Bell operator of Eq. ~(\ref{eq11}), the quantum upper bound is $3\sqrt{3} \approx 5.196$. On the other hand, we have proven in Sec. ~IID that  $3\sqrt{3} \cos{\frac{\pi}{18}} \approx 5.117 $ is the maximum value actually attainable, as given by Eq. ~(\ref{eq15}).

\section{Bell inequality for qudits}

We generalize the Bell inequality for qutrits to $d$-dimensional
systems, namely qudits, with $d$ a prime integer. A measurement on a
qudit produces one of $d$ possible outcomes. For a generalized Bell
inequality for qudits, two observers are allowed each to choose one of
$d$ variables. Consider a classical Bell function for qudits,
\begin{equation}
  \label{eq17}
  {\cal B}(\lambda) = \frac{1}{{d - 1}} \sum_{n = 1}^{d - 1} {\sum_{i = 0}^{d - 1} {\sum_{j = 0}^{d - 1} {\omega ^{n ij} A_i^n(\lambda) B_j^n(\lambda) } } },
\end{equation}
where $\omega$ is now a primitive $d$-th root of unity, i.e. $\omega =
\exp( i 2\pi/d)$, and $A_i(\lambda) = \omega ^{a_i(\lambda) }$ and
$B_j(\lambda) = \omega ^{b_j(\lambda) }$ with $a_i(\lambda)$ and
$b_j(\lambda)$ integer-valued functions of hidden variables $\lambda$.
Eq.~(\ref{eq17}) is reduced to the CHSH Bell function in Eq.~(\ref{e1})
if $d=2$ and to the two-qutrit function in Eq.~(\ref{eq4}) if $d=3$.
Similarly to the two-qutrit case, the Bell function in Eq.~(\ref{eq17})
can be rewritten as,
\begin{equation}
  \label{eq18}
  {\cal B}(\lambda) = \frac{d}{d - 1} \left( \sum_{i = 0}^{d - 1} \sum_{j = 0}^{d - 1} \delta ( a_i(\lambda)  + b_j(\lambda)  + ij ) - d \right),
\end{equation}
where $\delta(a) = 1$ if $a = 0 \bmod{d}$ and $\delta(a)=0$ otherwise.
As done in the two-qutrit case, we find classical upper and lower bounds
by considering $\Delta =\sum_{i,j} \delta ( a_i + b_j + ij)$. Using the
similar arguments as given from Eq.~(\ref{eq5}) to (\ref{eq9}), one
obtains $\Delta_{\mathrm{max}} = 3d-3$ and $\Delta_{\mathrm{min}}=0$.
Then, the statistical average of the Bell function satisfies the
following inequality,
\begin{equation}
  \label{eq24}
  - \frac{{d^2 }}{{d - 1}}\, \leq \langle {\cal B} \rangle\, \leq \frac{{d(2d - 3)}}{{d - 1}}
\end{equation}

The quantum Bell operator, corresponding to the classical Bell function,
is given as
\begin{equation}
  \label{eq25}
  \hat{\cal B} = \frac{1}{{d - 1}}\sum_{n = 1}^{d - 1} {\sum_{i = 0}^{d - 1} {\sum_{j = 0}^{d - 1} {\omega ^{nij} \hat A_i^n  \otimes \hat B_j^n } } },
\end{equation}
where $ \hat A_i$ and $\hat B_j$ are local unitary operators with
eigenvalues, $\{1,\omega, \omega^2, ..., \omega^{d-1}\}$. To show the
nonlocality, let the local operators be
\begin{equation}
  \label{eq26}
  \hat A_{j}  = \omega ^{j(j + 1)} \hat{f}_{1,j},\quad \hat B_{j} = \omega ^{\left( {\frac{{d + 1}}{2}} \right)^2 \left( {j^2  + 2j} \right)} \hat{f}_{1,\frac{{\left( {d + 1} \right)^2 }}{2}j}
\end{equation}
where $\hat f_{i,j}  = \hat X^i \hat Z^j$ and $\hat X$ and $\hat Z$ are now $d$-dimensional Pauli operators \cite{G99}. It is notable that $\hat{A}_i$ and $\hat{B}_j$ represent mutually unbiased measurements. Let further the two qudits be in a maximally entangled state,
\begin{equation}
  \label{eq27}
  | {\psi } \rangle  = \openone \otimes \hat{P} \frac{1}{\sqrt{d}}\sum_{k = 0}^{d - 1} | kk \rangle.
\end{equation}
where $\hat{P} = \sum_k \omega^{- \theta_k} |k \rangle\langle k|$. Here
$\theta_k$ is defined by
\begin{eqnarray}
  \frac{\theta_k}{d k} = \left\{
    \begin{array}{ll}
      \frac{d-1}{8} +
      \frac{(d+1)^2}{4 d k} \sum_{j=1}^k j^2, & \mbox{for $d=8m+1$} \\
      \frac{d+3}{8} - g_d +
      \frac{(d+1)^2}{4 d k} \sum_{j=1}^k j^2, & \mbox{for $d=8m+3$} \\
      \frac{d+3}{8} +
      \frac{(d+1)^2}{4 d k} \sum_{j=1}^k j^2, & \mbox{for $d=8m+5$} \\
      \frac{d-1}{8} -  g_d +
      \frac{(d+1)^2}{4 d k} \sum_{j=1}^k j^2, & \mbox{for $d=8m+7$} \\
    \end{array} \right.,
\end{eqnarray}
where $g_d = 0$ for $d = 8m+1$ or $8m+5$, and $g_d = 1/4d$ for $d =
8m+3$ or $8m+7$ for an integer $m$.  From the mutually unbiased 
local measurements of Eq.~(\ref{eq26}) and the maximally entangled state
in Eq.~(\ref{eq27}), the quantum expectation of the Bell operator is
given as
\begin{equation}
    \label{eq28}
    \langle {\psi } |\hat{\cal B} | {\psi}\rangle
    = \frac{1}{d(d - 1)}\sum_{n = 1}^{d - 1} \sum_{i,j,p = 0}^{d - 1}
    \omega ^{\xi( {i,j,p,n,g_d} ) },
\end{equation}
where $ \xi \left( {i,j,p,n,g_d} \right) = - 3 n g_d + n i j + \frac{{n(n - 1)}}{2}i + i n p + {\frac{3}{8} n (d - 1) + \frac{3}{d} \left( {\frac{{d + 1}}{2}} \right)^2 C(j,p,n)}$
and $ C ( {j,p,n} )=\sum_{k = 1}^n {( {j + p + k} )^2 } $. For $d=5$,
the quantum expectation is $25(1+ \sqrt{5})/8 \approx 10.113$. This is
clearly larger than the classical upper bound, $35/4 = 8.75$. For
$d=17$, the quantum expectation $\langle \psi |\hat{\cal B}| \psi
\rangle \approx 40.484$ exceeds 527/16=32.9375 of the classical upper
bound, while no violations are found for $d=7,11,13$ if local unitary
operators are employed as in Eq.~(\ref{eq26}).

Our Bell inequalities show relatively small degrees of violations. Ratios of
quantum to classical maxima are given for $d=3,5,17$ as:
\begin{eqnarray}
  \label{eq30}
  \frac{\langle \psi | \hat{\cal B} | \psi \rangle}{\langle {\cal B} \rangle} \approx
  \left\{
    \begin{array}{ll}
      1.137 &~~~~~\mbox{for $d=3$} \\
      1.156 &~~~~~\mbox{for $d=5$} \\
      1.229 &~~~~~\mbox{for $d=17$}
    \end{array}
  \right.. \nonumber
\end{eqnarray}
These ratios are smaller than 1.414 and 1.436, those of CHSH inequality
for qubits and CGLMP inequality for qutrits, respectively. However, it
is interesting to observe that the ratios increase with respect to the
dimension once the nonlocality appears.\\
\\

Let us now examine the robustness of our Bell inequality against the white noise. For this purpose, we consider the state
\begin{equation}
\label{add1}
\rho  = p\left| \psi  \right\rangle \left\langle \psi  \right| + \frac{{\left( {1 - p} \right)}}{{d^2 }} \openone \otimes \openone
\end{equation}
This state represents a mixture of the pure state of Eq. ~(\ref{eq27}) and the fully mixed state, where $p$ is the relative weight of the pure state $ \left| \psi \right\rangle $ with respect to the fully mixed state. We compute the lower bound $p_{min}$ of the $p$ value above which our Bell inequality is violated. Our calculation shows that $p_{min}=0.88,\;0.8653$ and $0.814$ for $d=3,\;5$ and $17$, respectively. One thus sees that our Bell inequality is more robust against the white noise as the dimension $d$ is increased, the tendency also observed in the CGLMP inequality. 

\section{Summary}

We proposed a Bell inequality for two qutrits. This Bell inequality is
maximally violated by quantum mechanics for mutually unbiased
measurements and a maximally entangled state, whereas other Bell
inequalities for high-dimensional systems such as CGLMP and that of Son {\it et al.} do not
satisfy those conditions. This feature is consistent with the CHSH inequality of two qubits. Note that our Bell inequality consists of three settings of
local measurements while CHSH, CGLMP and the inequality of Son {\it et al.} have two settings.

The Bell inequality for qutrits was generalized to prime-dimensional
qudits. We investigated the generalized Bell inequalities for two qudits
with the dimensions up to 17, finding the nonlocality for the dimensions
$5$ and $17$. Further studies on the generalized Bell inequalities are
encouraged to clarify if there are violations for higher dimensional
systems and if the degree of nonlocality persistently increases with
respect to the dimension once the nonlocality appears.
\section*{Acknowledgments}

SWJ and HWL were supported by a Grant from Korea Research Institute for
Standards and Science (KRISS). JL was supported by the Korean Research Foundation Grant funded by the
Korean Government (MOEHRD) (KRF-2005-041-C00197). KN was supported by Frontier Basic Research Program at KAIST and by a BK21 research grant. The authors thank Professor M. S. Kim of Queen's University, Belfast
for helpful discussions.

\end{document}